\documentclass[a4paper,twoside,prd,nofootinbib,preprintnumbers,twocolumn,floatfix]{revtex4}
\usepackage{amssymb,amsmath,bm,natbib}
\usepackage{color}
\usepackage{slashed}
\usepackage{graphics}
\usepackage{graphicx}
\usepackage[utf8]{inputenc}
\usepackage{tabularx}
\usepackage{adjustbox}
\usepackage{siunitx}
\usepackage[caption=false]{subfig}
\usepackage{hyperref}
\usepackage{url}
\usepackage{dsfont}
\usepackage{cancel}
\usepackage{float}
\usepackage{multirow}
\usepackage{booktabs}
\usepackage{array}
\usepackage{pdflscape}
\usepackage{wrapfig}
\usepackage{csquotes}
\usepackage{soul}

\newcommand{\Cc}[1]{{\cal C}_{#1}}

\newcommand{\TeV}{{\rm TeV}}

\newcommand{\eq}[1]{Eq.~\eqref{#1}}

\newcommand{\C}[1]{{\cal C}_{#1}}

\begin{document}
\preprint{ZU-TH 21/22, PSI-PR-22-14 }
\title{Disentangling Lepton Flavour Universal and \\Lepton Flavour Universality Violating Effects in $b\to s\ell^+\ell^-$ Transitions}

\author{Marcel Alguer\'{o}}
\email{malguero@ifae.es}
\affiliation{Grup de Fisica   Te\`orica (Departament de Fisica), Universitat Aut\`onoma de Barcelona, E-08193 Bellaterra (Barcelona)}
\affiliation{Institut de Fisica d'Altes Energies (IFAE),
	The Barcelona Institute of Science and Technology, Campus UAB, E-08193 Bellaterra (Barcelona)}

\author{Bernat Capdevila}
\email{bernat.capdevilasoler@unito.it}
\affiliation{Università di Torino and INFN Sezione di Torino, Via P. Giuria 1, Torino I-10125, Italy}

\author{Andreas Crivellin}
\email{andreas.crivellin@cern.ch}
\affiliation{Paul Scherrer Institut, CH--5232 Villigen PSI, Switzerland}
\affiliation{Physik-Institut, Universit\"at Z\"urich, Winterthurerstrasse 190, CH--8057 Z\"urich, Switzerland  }

\author{Joaquim Matias}
\email{matias@ifae.es}
\affiliation{Grup de Fisica   Te\`orica (Departament de Fisica), Universitat Aut\`onoma de Barcelona, E-08193 Bellaterra (Barcelona)}
\affiliation{Institut de Fisica d'Altes Energies (IFAE),
	The Barcelona Institute of Science and Technology, Campus UAB, E-08193 Bellaterra (Barcelona)}

\begin{abstract}
In this letter we propose a strategy for discerning if new physics in the Wilson coefficient ${\cal C}_{9\mu}$ is dominantly lepton flavour universality violating  or if it contains a sizable  lepton flavour universal component (${\cal C}_9^{\rm U}$).
Distinguishing among these two cases, for which the model independent fit of the related scenarios exhibits similar pulls with respect to the Standard Model, is crucial to advance our understanding of the $B$ anomalies. We first identify the origin of the degeneracy of these two cases and point out the key observables that can break it. In particular, while the observables  measured so far that test lepton flavour universality exhibit similar dependencies to all the relevant  Wilson coefficients, the forthcoming measurement of $Q_5=P_{5}^{\prime\mu}-P_{5}^{\prime e}$ is particularly sensitive to ${\cal C}_{9\mu}-{\cal  C}_{9e}$. In fact, if $Q_5$ were found to be small (i.e.~close to its Standard Model value), 
this would imply  a small ${\cal C}_{9\mu}-{\cal  C}_{9e}$ but a sizable ${\cal C}_{9}^{\rm U}$, given the preference of global fits for a large negative new physics contribution in ${\cal C}_{9\mu}$.  We discuss the possible origins of ${\cal C}_9^{\rm U}$, in particular how it could originate from new physics. Here, a promising scenario, that could even link $b\to s \ell^+\ell^-$ to $R_{D^{(*)}}$, is the one in which ${\cal C}_9^{\rm U}$ is generated from a tau loop via an off-shell photon penguin  diagram. This setup predicts the branching ratios of $B_s\to\tau^+\tau^-$ and $B\to K^{(*)}\tau^+\tau^-$ to lie within the reach of LHCb, CMS and Belle~II. Alternatively, in case of a non-observation of these tauonic processes, we show that the most natural possibility to generate ${\cal C}_9^{\rm U}$ is a $Z^\prime$ with partially lepton flavour universal couplings.  \bigskip

\end{abstract}
\maketitle

\section{Introduction and Motivation}
\vspace*{-0.4cm}
In $b\to s \ell^+\ell^-$ transitions, intriguing hints for physics beyond the Standard Model (SM) have been uncovered. In particular, the accumulated evidence for new physics (NP) obtained by combining many different channels points convincingly towards a breakdown of the SM at the (multi) TeV scale. These global fits are, in principle, capable of identifying which specific NP patterns describe data best. However, it was found in Refs.~\cite{Alguero:2021anc,Hurth:2021nsi} that several different NP patterns (or scenarios) are preferred over the SM hypothesis with a similar significance of more than $7\sigma$ (see also Refs.~\cite{Altmannshofer:2021qrr,Kowalska:2019ley,Blake:2019guk,Ciuchini:2017mik}). In fact, since the appearance of the first anomalies in the angular observables, most prominently $P_5^{\prime\mu}$ in $B \to K^* \mu^+\mu^-$~\cite{Descotes-Genon:2012isb,LHCb:2020lmf}, and later on in the ratios $R_{K^{(*)}}$~\cite{Hiller:2003js,LHCb:2021trn,LHCb:2017avl,LHCb:2021lvy}, a main goal of the model-independent global fits was to identify which pattern of NP can account best for data. However, with each new update (see Ref.~\cite{London:2021lfn} for details  and 
Ref.~\cite{Alguero:2021yus} for future progress) the pull with respect to the SM for the already preferred scenarios increased while no clear discrimination among them arose. In this context, the Wilson coefficient ${\cal C}_{9\mu}$ of the operator $\left(\bar{s}\gamma_\mu P_L b\right)\left(\bar{\mu} \gamma^\mu \mu \right)$ plays the prominent role because its NP contribution is sizable in all scenarios that describe data well. Nonetheless, it is not yet clear if the NP contribution entering ${\cal C}_{9\mu}$ is dominantly related to muons, i.e.~if it has a large lepton flavour universality violating (LFUV) component, or if it contains a sizable lepton flavour universal (LFU) term. 

In fact, all patterns that describe data well can be basically classified in two groups depending on the size of the LFUV component of ${\cal C}_{9\mu}$ (or inversely of the LFU one). Therefore, distinguishing among these two possibilities (small or sizable LFUV component) is crucial to advance our understanding of which model could lie behind the $B$ anomalies. However, while refining the measurements of the ratios $R_{X}$ (with $X=K,K^*,K_S,K^{*+},\phi$) is of utmost importance to establish if LFU is violated in Nature, it will not help in the near future in making progress towards disentangling between these two cases. 

In this paper we will show that the inclusion of a different kind of LFUV observable (other than the $R_X$ ones) is decisive to break the degeneracy among the scenarios with large and small LFUV component in $\Cc{9\mu}$ to finally make progress in the identification of the pattern of NP realized in Nature. Indeed, the use of a $\Cc{9}$-dominated observable is a more promising strategy in the short term to answer this question, rather than collecting more and more statistics for $R_X$-type observables\footnote{Nonetheless, it is clear that both strategies (increasing statistics and targeting specific observables) should be implemented in the experimental program.}.

Currently, the best candidate for a $\Cc{9}$-dominated observable testing LFU that can discriminate among the leading scenarios is a measurement of $Q_5=P_{5}^{\prime\mu}-P_{5}^{\prime e}$\cite{Capdevila:2016ivx}. We will lay out a path based on this observable and $b \to s \tau^+\tau^-$ processes that can be implemented in the near future to clarify the composition of $\Cc{9\mu}$. In particular, the proposed strategy will help identifying the nature of the NP contributions entering ${\cal C}_{9\mu}$, i.e.~how big the LFU component is, using the plausible direct link with $b\to s\tau^+\tau^-$ processes such as ${\cal B}_{B \to K^{(*)}\tau^+\tau^-}$ and ${\cal B}_{B_s \to \tau^+\tau^-}$~\cite{Capdevila:2017iqn,Crivellin:2018yvo}, motivated by the $R_{D^{(*)}}$ anomalies.  As a byproduct we will also be able to quantify if some marginal, so far unknown, extra hadronic contribution affects $\Cc{9\mu}$. 

The structure of this paper is as follows.
In Sec.~\ref{sec:setup} we discuss the anatomy of the Wilson coefficient $\Cc{9\mu}$. Sec.~\ref{sec:Disentangling} describes how to determine the LFUV and LFU piece of this 
coefficient, while Sec.~\ref{sec:scenarios} discusses the possible origins of the LFU component. Finally, Sec.~\ref{sec:NPQ5} shows that $Q_5$ classifies scenarios in two groups to conclude in Sec.~\ref{sec:conclusions}.

\section{Setting the stage}
\label{sec:setup}

The weak effective Hamiltonian relevant to the processes considered here is~\cite{Buchalla:1995vs}
\begin{equation}
{\mathcal H}_{\rm eff}=-\frac{4G_F}{\sqrt{2}} V_{tb}V_{ts}^*\sum_i \C{i}  {\mathcal O}_i + {\rm h.c.}\,,
\end{equation}
where the semileptonic operators we will focus on are (see Ref.~\cite{Chetyrkin:1996vx,Bobeth:1999mk,Capdevila:2017bsm} for details and the definition of the rest of operators)
\begin{eqnarray}
{\mathcal{O}}_{9\ell} &=& \frac{e^2}{16 \pi^2} 
(\bar{s} \gamma_{\mu} P_L b)(\bar{\ell} \gamma^\mu \ell) \,,\\
{\mathcal{O}}_{{9^\prime\ell}} &=& \frac{e^2}{16 \pi^2} 
(\bar{s} \gamma_{\mu} P_R b)(\bar{\ell} \gamma^\mu \ell) \,, \\
{\mathcal{O}}_{10\ell} &=&\frac{e^2}{16 \pi^2}
(\bar{s}  \gamma_{\mu} P_L b)(  \bar{\ell} \gamma^\mu \gamma_5 \ell) \,,\\
{\mathcal{O}}_{{10^\prime\ell}} &=&\frac{e^2}{16\pi^2}
(\bar{s}  \gamma_{\mu} P_R b)(  \bar{\ell} \gamma^\mu \gamma_5 \ell) \,.
\end{eqnarray}

The Wilson coefficient $\Cc{9\ell}$, that plays a prominent role in this discussion, has a rather complicated structure compared to the other Wilson coefficients. On the one hand, it always appears in observables in combination with the Wilson coefficients of the four-quark operators ${\cal O}_{1,..,6}$ (with ${\cal O}_i=4 P_i$ and $P_i$ defined in Ref.~\cite{Chetyrkin:1996vx}) and hence an ``effective'' superscript is included to denote this combination (see Ref.~\cite{Buras:1993xp}). On the other hand, the non-local contributions from $c\bar{c}$ loops enter the same amplitude structures as ${\cal  O}_9$ (and ${\cal O}_7$). Therefore, they can be naturally recast as a $q^2$ (dilepton invariant mass), helicity and process dependent contribution accompanying the perturbative SM term $\Cc{9\mu \, \rm pert}^{\rm SM}$. 

We thus write (for the $B\to K^*\ell^+\ell^-$ case)
\begin{equation}
\label{decomp}
{\cal C}_{9\mu \, j}^{{{\rm eff} \, B \to K^*} 
}={\cal C}_{9\mu \, \rm pert}^{\rm SM}+ {\cal C}_{9\mu}^{\rm NP} + {\cal C}_{9\mu \, j}^{c\bar{c} \, B \to K^*}(q^2)  \,,
\end{equation}
where ${\cal C}_{9\mu \, j}^{c\bar{c} \, B \to K^*}=s_j {\cal C}_{9\mu \, j \, \rm KMPW}^{c\bar{c} \, B \to K^*}$ stands for the one soft-gluon non-factorizable leading long-distance $c\bar{c}$ contribution. 
This contribution is transversity amplitude dependent, therefore we introduce the subscript $j$ for each amplitude $j=\perp,||,0$~\cite{Khodjamirian:2010vf} of the $B\to K^*\ell^+\ell^-$ decay\footnote{For the case of a $B \to K$ transition, the $j$ subscript should be removed and the corresponding ${\cal C}_{9\mu  \, \rm KMPW}^{c\bar{c} \, B \to K}$~\cite{Khodjamirian:2012rm} should be used.}. Finally, a $s_j$ nuisance parameter is included  and allowed to vary  from -1 to 1  for each amplitude independently due to the theoretical difficulty to ascertain the relative phase between the long and short distance contributions and in order to be very conservative. For a discussion at length on our treatment of the charm-loop contributions we refer the reader to Refs.~\cite{Descotes-Genon:2014uoa,Capdevila:2017ert}.

A first step in the analysis of $\Cc{9\mu}^{\rm NP}$, in order to fix our notation for the next sections, is to split the NP piece 
of $\Cc{9\mu}^{\rm NP}$ into two:
\begin{eqnarray} 
\label{c9np}
{\cal C}_{9\mu}^{\rm NP}=  {\cal C}_{9\mu}^{\rm V} + {\cal C}_9^{\rm U}\,,\qquad {\cal C}_{9e}^{\rm NP}=  {\cal C}_9^{\rm U}\,, 
\end{eqnarray}
where ${\rm V}$ $({\rm U})$ stands for the LFUV (LFU) contribution, respectively. Finally, if some hypothetical additional hadronic effect beyond the long-distance charm one already included\footnote{In recent years there has been an important progress on the theoretical side to evaluate more precisely the impact of long-distance charm contribution entering ${\cal C}_9^{\rm eff}$ either using explicit computations or resonance data to model it~\cite{Bobeth:2017vxj}. Particularly important is the result presented in Ref.~\cite{Gubernari:2020eft} where it was found that the NLO contribution is almost negligible as compared to the LO soft-gluon exchange computed in Ref.~\cite{Khodjamirian:2010vf}.} exists, it would in general be process and $q^2$ dependent, like the long-distance charm contribution, and manifest itself in inconsistent values for $\Cc{9\mu}^{\rm NP}$ when comparing its determination from different channels\footnote{For instance, the long-distance charm contribution entering $B \to K \ell^+\ell^-$ was found to be  negligible in Ref.~\cite{Khodjamirian:2012rm} compared to the one of $B\to K^*\ell^+\ell^-$~\cite{Khodjamirian:2010vf}. Furthermore, with the present precision, the different channels within the global fits points to a consistent picture, disfavouring such a hypothesis.}. However, if it were, for unknown reasons, $q^2$, process and helicity independent, such a hypothetical contribution could hide inside $\Cc{9}^{\rm U}$ mimicking a universal NP contribution.

In the following sections we will propose a strategy based on identifying specific observables (i.e.~$Q_5$ and $b\to s\tau^+\tau^-$ processes) able to disentangle the LFUV from the LFU piece in Eq.~\eqref{c9np} and bound the possible size of extra ``universal" hadronic contributions. In turn, the addition of those observables in the global fits will break present degeneracies among scenarios with large and small $\Cc{9\mu}^{\rm V}$, providing a guideline to identify the NP pattern realized in Nature.

\bigskip
\begin{figure*}[t] 
    \centering
    \includegraphics[width=0.44\textwidth]{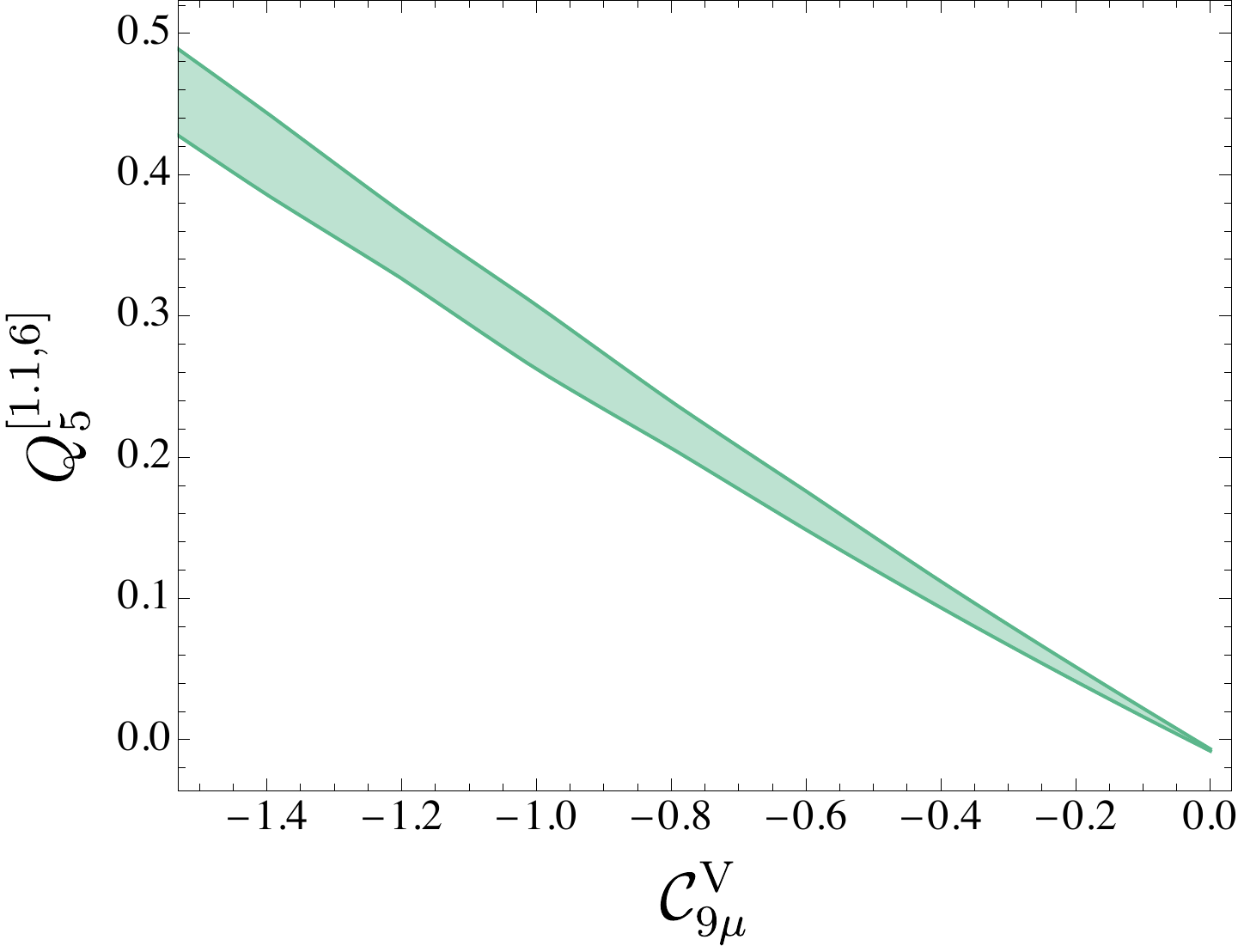}~~~~
        \includegraphics[width=0.44\textwidth]{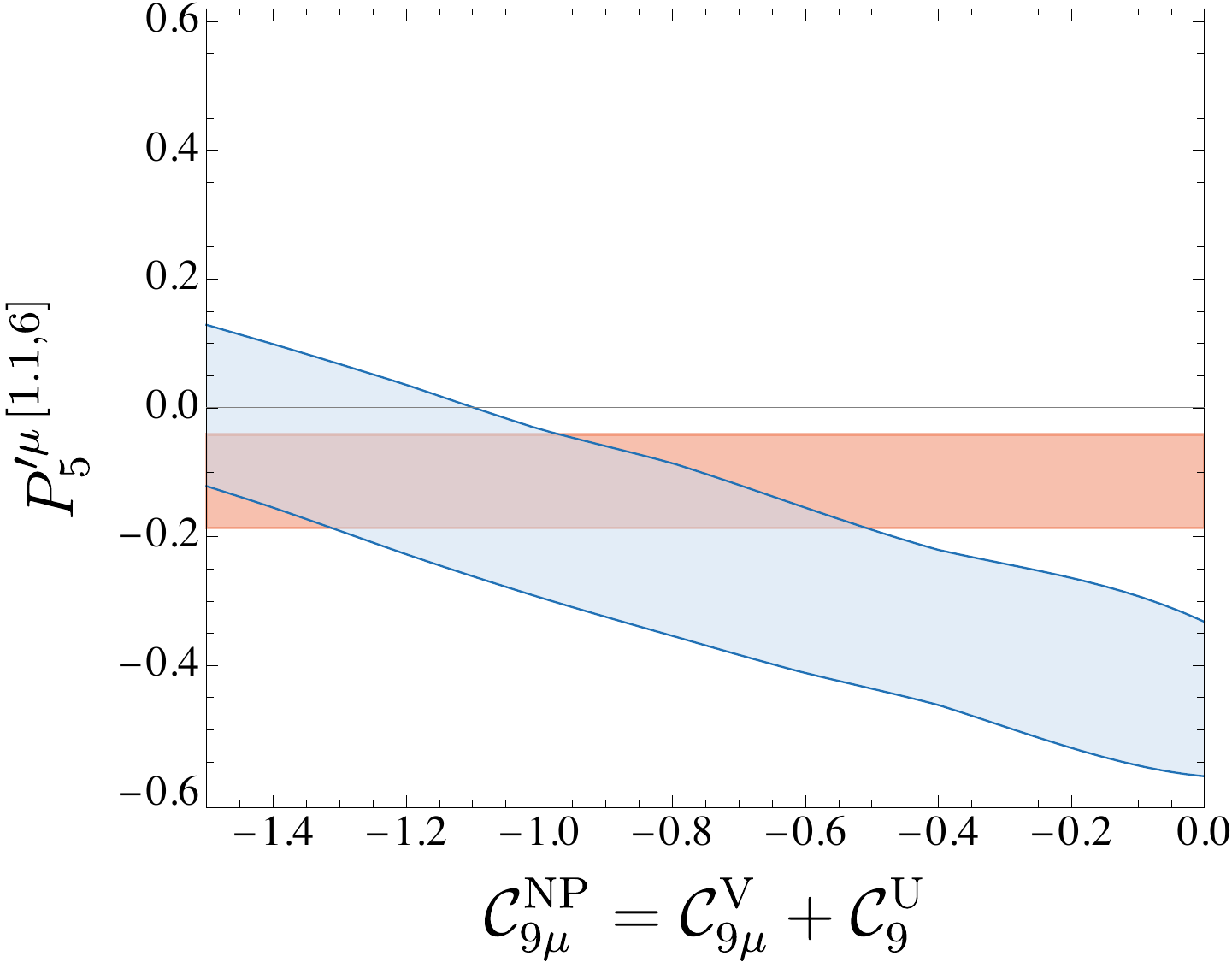}
    \caption{Left: Prediction of $Q^{[1.1,6]}_5$ as a function of $\Cc{9\mu}^{\rm V}$. Right: Prediction for $P^{\prime \mu \, [1.1,6]}_5$  as a function of $\Cc{9\mu}^{\rm NP}$ (blue band). The current experimental value from the LHCb collaboration~\cite{LHCb:2020lmf} is indicated as an orange horizontal band. }
    \label{fig:Q5bin16}
\end{figure*}

\section{Disentangling LFU from LFUV effects in \texorpdfstring{${\cal C}_{9\mu}$}{C9}}
\label{sec:Disentangling}
\subsection{Determining \texorpdfstring{${\cal C}_{9\mu}^{\rm V}$}{C9V}}

First of all, note that a non-zero $\Cc{9\mu}^{\rm V}$ must undoubtedly be of NP origin given that the SM gauge interactions do not discriminate between the lepton families. Since due to the measurements of $R_K$ and $R_{K^*}$ all scenarios that allow for a very good fit to data require of $\Cc{9\mu}^{\rm V}$  to some degree, it is clear that also $P_5^{\prime\mu}$ (or other $b\to s \mu^+\mu^-$ observables) must necessarily contain a NP contribution. But how can one determine the size of ${\cal C}_{9\mu}^{\rm V}$? For this purpose one should focus on observables that I) can disentangle the LFUV piece from the LFU one and II) are ${\cal C}_9$-dominated. Condition I is fulfilled by all observables measuring LFUV, however, not all LFUV observables are equally instrumental (or useful) in separating the two pieces (i.e.~do not fulfill condition II) because of the different relative weights of the Wilson coefficients entering the LFUV observables. 

In order to illustrate this, we have computed the expressions for $R_K$, $R_{K^*}$ as semi-analytic functions of the Wilson coefficients in the [1.1,6]~GeV$^2$ bin
{\small
\begin{eqnarray} \label{rk}
R_K&\!\!\!=\!\!\!&[1 -0.26 (\Cc{10\mu}^{\rm NP}+\Cc{10'\mu})+0.23 (\Cc{9\mu}^{\rm NP}+\Cc{9'\mu})+S^{K}_\mu]/ \nonumber \\
&&[1 -0.26 (\Cc{10e}^{\rm NP}+\Cc{10'e})+0.23 (\Cc{9e}^{\rm NP}+\Cc{9'e})+S^{K}_e]\,, \\
R_{K^*}&\!\!\!=\!\!\!&[1 -0.29 \Cc{10\mu}^{\rm NP}+ 0.20 \Cc{10'\mu} + 0.16 (\Cc{9\mu}^{\rm NP} -\Cc{9'\mu})+S^{K^*}_\mu] /
\nonumber \\
&&[1 -0.29 \Cc{10e}^{\rm NP}+ 0.20 \Cc{10'e} + 0.16 (\Cc{9e}^{\rm NP} - \Cc{9'e})+S^{K^*}_e]\,,\nonumber
\end{eqnarray}
}
\noindent
where $S^K_\ell=0.03\, [ (\Cc{9\ell}^{\rm NP}+\Cc{9'\ell})^2+(\Cc{10\ell}^{\rm NP}+\Cc{10'\ell})^2]$ and $S^{K^*}_\ell=0.03\, [ (\Cc{9\ell}^{\rm NP}-\Cc{9'\ell})^2+(\Cc{10\ell}^{\rm NP}-\Cc{10'\ell})^2]+0.02\,[\Cc{9\ell}^{\rm NP}\Cc{9'\ell}+\Cc{10\ell}^{\rm NP}\Cc{10'\ell}]$ are subleading quadratic terms\footnote{Note that all the semianalytic expressions presented in this paper should be taken as a guideline to understand the sensitivity to NP that an observable exhibits, neglecting uncertainties. However, these uncertainties should always be included when determining the Wilson coefficients (see Fig.~\ref{fig:Q5bin16}).}.
Expanding the observables in Eq.~\eqref{rk}  in the limit of small values for the NP Wilson coefficients naturally give rise to an expression in terms of  $\Cc{i\mu}^{\rm V}=\Cc{i \mu}-\Cc{i e}$ (analogous to Eq.\eqref{c9np})  and subleading crossed terms $\Cc{i\mu}^{\rm V} \Cc{j}^{\rm U}$.
These expressions show that while these observables are fundamental to establish the presence of LFUV NP, they are not particularly useful to pin down one among the different currently preferred scenarios. This is clear from the expression of $R_K$, where $\Cc{9\mu}^{\rm NP}$ and $\Cc{10\mu}^{\rm NP}$ have approximately the same weight, such that e.g.~$\Cc{9\mu}^{\rm NP}=-1$ results in nearly the same prediction as $\Cc{9\mu}^{\rm NP}=-\Cc{10\mu}^{\rm NP}=-0.5$. Furthermore, even though in the case of $R_{K^*}$ the sensitivity to $\Cc{9\mu}^{\rm NP}$ and $\Cc{10\mu}^{\rm NP}$ differs by near a factor of two, this is not enough to determine ${\cal C}_{9\mu}^{\rm V}$ precisely due to the hadronic uncertainties affecting $R_{K^{(*)}}$ in presence of NP. 

These considerations clearly imply that we cannot discern from the $R_{K^{(*)}}$ observables $\Cc{9\mu}^{\rm V}$ due to the similar dependence on $\Cc{10\mu}^{\rm V}$ (and also on primed Wilson coefficients) that they exhibit; i.e.~the $R_{K^{(*)}}$ ratios do not fulfill condition II. Instead, a LFUV observable that fulfills condition II is $Q_5$ (evaluated on the same bin [1.1,6] GeV$^2$):
\begin{eqnarray}
Q_5&=&Q_5^{\rm SM}-0.25  \Cc{9\mu}^{\rm V} + 0.20  \Cc{10'\mu}^{\rm V}+S^{Q5}\,,
\label{q5}
\end{eqnarray}
with $Q_5^{\rm SM}=-0.01$ and 
where, 
contrary to $R_X$ observables, the subleading contribution includes also linear terms
\begin{eqnarray} 
S^{Q_5}&=&-0.02  \Cc{10\mu}^{\rm V} 
-0.04  \Cc{9'\mu}^{\rm V} +
0.03  {\Cc{9\mu}^{\rm V}}^2 +0.07  \Cc{9\mu}^{\rm V} \Cc{9}^{\rm U} \nonumber \\
&&
 +0.05 [  \Cc{10\mu}^{\rm V}  \Cc{10'\mu}^{\rm V} +  \Cc{10\mu}^{\rm V} \Cc{10^\prime}^{\rm U} +  \Cc{10^\prime\mu}^{\rm V} \Cc{10}^{\rm U}]   \nonumber \\
&& 
-0.03 [  \Cc{9\mu}^{\rm V}  \Cc{9'\mu}^{\rm V} +  \Cc{9\mu}^{\rm V} \Cc{9^\prime}^{\rm U} +  \Cc{9^\prime\mu}^{\rm V} \Cc{9}^{\rm U} + \Cc{10^\prime\mu}^{\rm V}  \Cc{10'}^{\rm U}]\, \nonumber \\ 
 && -0.04 [  \Cc{9'\mu}^{\rm V}  \Cc{10'\mu}^{\rm V} +  \Cc{10'\mu}^{\rm V} \Cc{9^\prime}^{\rm U} +  \Cc{9^\prime\mu}^{\rm V} \Cc{10^\prime}^{\rm U}], 
 \end{eqnarray}
where the last three terms correspond to small cross products of LFUV and LFU contributions. Eq.~\eqref{q5} shows that $Q_5$, contrary to $R_{K^{(*)}}$, breaks the approximate degeneracy of the two scenarios ($\Cc{9\mu}^{\rm NP}$ versus $\Cc{9\mu}^{\rm NP}=-\Cc{10\mu}^{\rm NP}$) as the dominant contributions in $Q_5$ originate from $\Cc{9\mu}^{\rm V}$ and $\Cc{10^\prime\mu}^{\rm V}$ (or combined as $\tilde{\cal C}_{9\mu}^{\rm V}=\Cc{9\mu}^{\rm V}-0.8\Cc{10^\prime\mu}^{\rm V}$) with $\Cc{10\mu}^{\rm V}$ playing a negligible role as can be seen from $S^{Q_5}$. Moreover, the structure of $Q_5$ and the preference of the global fits~\cite{Alguero:2021anc} for a large negative $\Cc{9\mu}^{\rm V}$ with a small (also) negative $\Cc{10^\prime\mu}^{\rm V}$, implies that a measurement of $Q_5$, even in presence of right-handed currents (RHCs), provides a lower bound on the absolute value of $|\Cc{9\mu}^{\rm V}|$. For instance, the best fit point of the 1D scenario with only $\Cc{9\mu}^{\rm NP}$ is $\Cc{9\mu}^{\rm NP}=-1.01$, while in the 2D scenario with $\Cc{9\mu}^{\rm NP}$ and $\Cc{10^\prime\mu}$  it was found~\cite{Alguero:2021anc} $\{\Cc{9\mu}^{\rm NP}=-1.15,\Cc{10^\prime\mu}=-0.26\}$. Therefore, 
\begin{equation}
\label{c9v}
|\Cc{9\mu}^{\rm V}| = |\tilde{{\cal C}}_{9\mu}^{\rm V} +0.8 {\cal C}_{10^\prime\mu}^{\rm V}| 
\geq |\tilde{\cal C}_{9\mu}^{\rm V}|\,,
\end{equation}
where $\tilde{{\cal C}}_{9\mu}^{\rm  V}$ is obtained using
\begin{equation}
\label{q5aprox}
Q_5\simeq Q_5^{\rm SM}-0.25  \tilde{{\cal C}}_{9\mu}^{\rm V}\,.
\end{equation}
The left plot in Fig.~\ref{fig:Q5bin16} shows $Q_5$ versus $\Cc{9\mu}^{\rm V}$, i.e.~the value of $\Cc{9\mu}^{\rm V}$ that can be inferred from a measurement of $Q_5$ in absence of any other source of NP. 

In summary, even in presence of RHCs contributing to $\Cc{10^\prime\mu}^{\rm V}$, $Q_5$ provides a lower bound (or a measurement in absence of them) on ${\cal C}_{9\mu}^{\rm V}$. Note that in the future a precise measurement of the observables $P_1^\ell$ (with $\ell=e,\mu$)~\cite{Kruger:2005ep,Lunghi:2006hc,Descotes-Genon:2013vna} can help to identify the presence of RHCs.  This observable is approximately zero in the large-recoil region in absence of RHCs. Therefore a deviation observed in $P_1^\ell$ will indicate non-zero primed Wilson coefficients.

\subsection{Determining \texorpdfstring{${\cal C}_{9}^{\rm U}$}{C9U}}

The next step is to identify the LFU piece $\Cc{9}^{\rm U}$ using $P_5^{\prime\mu}$ and in the future also $P_5^{\prime e}$. For this we calculate $P_5^{\prime \mu}$ in the [1.1,6]~GeV$^2$ bin which follows from $Q_5$ with obvious substitutions:
\begin{eqnarray}
\label{p5}
P_5^{\prime \mu}&=&P_{5}^{\prime \mu \, {\rm SM}}-0.25  \Cc{9\mu}^{\rm NP} + 0.20  \Cc{10'\mu}+ S^{P_5^{\prime\mu}}\,,
\end{eqnarray}
where $P_{5}^{\prime \mu \, {\rm SM}}=-0.44$ and 
\begin{eqnarray}
S^{P_5^{\prime\mu}}&=&-0.02\Cc{10\mu}^{\rm NP}-0.04\Cc{9^\prime\mu}+0.03\Cc{9\mu}^{\rm NP2}-0.03\Cc{9\mu}^{\rm NP} \Cc{9^\prime\mu} \nonumber \\
 &&+0.05 \Cc{10\mu}^{\rm NP} \Cc{10^\prime\mu} -0.04  \Cc{9^\prime\mu}^{\rm NP} \Cc{10^\prime\mu}.
\end{eqnarray}
Our central point here is to show that one can determine a lower bound on $\Cc{9}^{\rm U}$ as well using a similar strategy as for $Q_5$. Eq.~\eqref{p5} can be approximated to
\begin{equation}
P_5^{\prime \mu}\simeq P_{5}^{\prime \mu \, {\rm SM}}-0.25  \tilde{\cal C}_{9\mu}\,,
\label{p5paprox}
\end{equation}
where $\tilde{\cal C}_{9\mu}=\Cc{9\mu}^{\rm NP}-0.8 \Cc{10^\prime\mu}$. Then if we have data available only from $Q_5$ and $P_5^{\prime \mu}$ we obtain immediately an upper bound on the universal piece from\footnote{ In the same way, we can obtain, using the definitions of $\tilde{\cal C}_{9\mu}$ and $\tilde{\cal C}_{9\mu}^{\rm V}$, not just a bound, but the total amount of LFU from $\tilde{\cal C}_{9\mu}-\tilde{\cal C}_{9\mu}^{\rm V}={\cal C}_{9}^{\rm U}-0.8 {\cal C}_{10^\prime}^{\rm U}$.}:
\begin{equation} 
\label{c9utot}
 |{\cal C}_{9}^{\rm U}| = |\tilde{\cal C}_{9\mu}-\tilde{\cal C}_{9\mu}^{\rm V}+0.8 {\cal C}_{10^\prime}^{\rm U}| \geq |\tilde{\cal C}_{9\mu}-\tilde{\cal C}_{9\mu}^{\rm V}|\,,
\end{equation}
where the terms on the r.h.s.~correspond to the measurement of the NP piece entering $P^{\prime \mu}_{5}$ and $Q_5$ using Eq.~\eqref{q5aprox} and Eq.~\eqref{p5paprox}, respectively. Notice that in the absence of universal RHC contributions the bound turns into a measurement. Instead, if experiments also provided data on the electronic angular observable $P_5^{\prime e}$ we could directly get
\begin{equation} \label{c9ue}
|{\cal C}_{9}^{\rm U}|=|\tilde{\cal C}_{9e}+0.8 {\cal C}_{10^\prime}^{\rm U }| \geq |\tilde{\cal C}_{9 e}|\,,
\end{equation}
where the Wilson coefficients for electrons are obtained from Eq.~\eqref{p5paprox} by replacing $\mu$ with $e$, where a preference from global fits for a negative sign in both coefficients ($\Cc{9}^{\rm U}$ and $\Cc{10^\prime}^{\rm U}$) is also required (see scenario 7 and 11 in Ref.\cite{Alguero:2021anc}). 

Note that in contrast to $\Cc{9\mu}^{\rm V}$, which can only originate from NP, the LFU piece, as pointed out earlier could in principle contain, together with the NP contribution, some residual higher-order hadronic effect. For this reason we have to proceed differently for the LFU piece, identifying and measuring separately all possible sources of LFU NP. The difference between $\Cc{9}^{\rm U}$ obtained following Eq.~\eqref{c9utot} and Eq.~\eqref{c9ue}, and the one obtained from the different NP sources (called from now on $\Cc{9}^{\rm NP \, U}$), provides the best strategy to identify the actual amount (if any) of an unknown hadronic contribution.

\section{Possible origins of \texorpdfstring{${\cal C}_{9}^{\rm U}$}{C9U}}
\label{sec:scenarios}

Assuming $Q_5$ is found to be small, this would imply also a small ${\cal C}_{9\mu}^{\rm V}$ but a sizeable ${\cal C}_{9}^{\rm U}$, given the present preference of global fits for a large negative ${\cal C}^{\rm NP}_{9\mu}$. Then the question of a possible origin of the LFU contribution arises since most models considered in the literature generate purely LFUV effects (see Refs.~\cite{Crivellin:2019dun,Allanach:2019iiy,Crivellin:2020oup,Alguero:2022est} for some exceptions). There are three (non-exclusive) possibilities how one could obtain an effect in ${\cal C}_{9}^{\rm U}$. First of all, it could originate, as discussed above, from some unknown and intricate hadronic effect. Second, it could be generated via a direct LFU NP contribution, and finally it could be induced indirectly from a NP effect via renormalization group effects originating from loops involving light SM particles.

\begin{figure*}
\centering
    \includegraphics[width=0.46\textwidth]{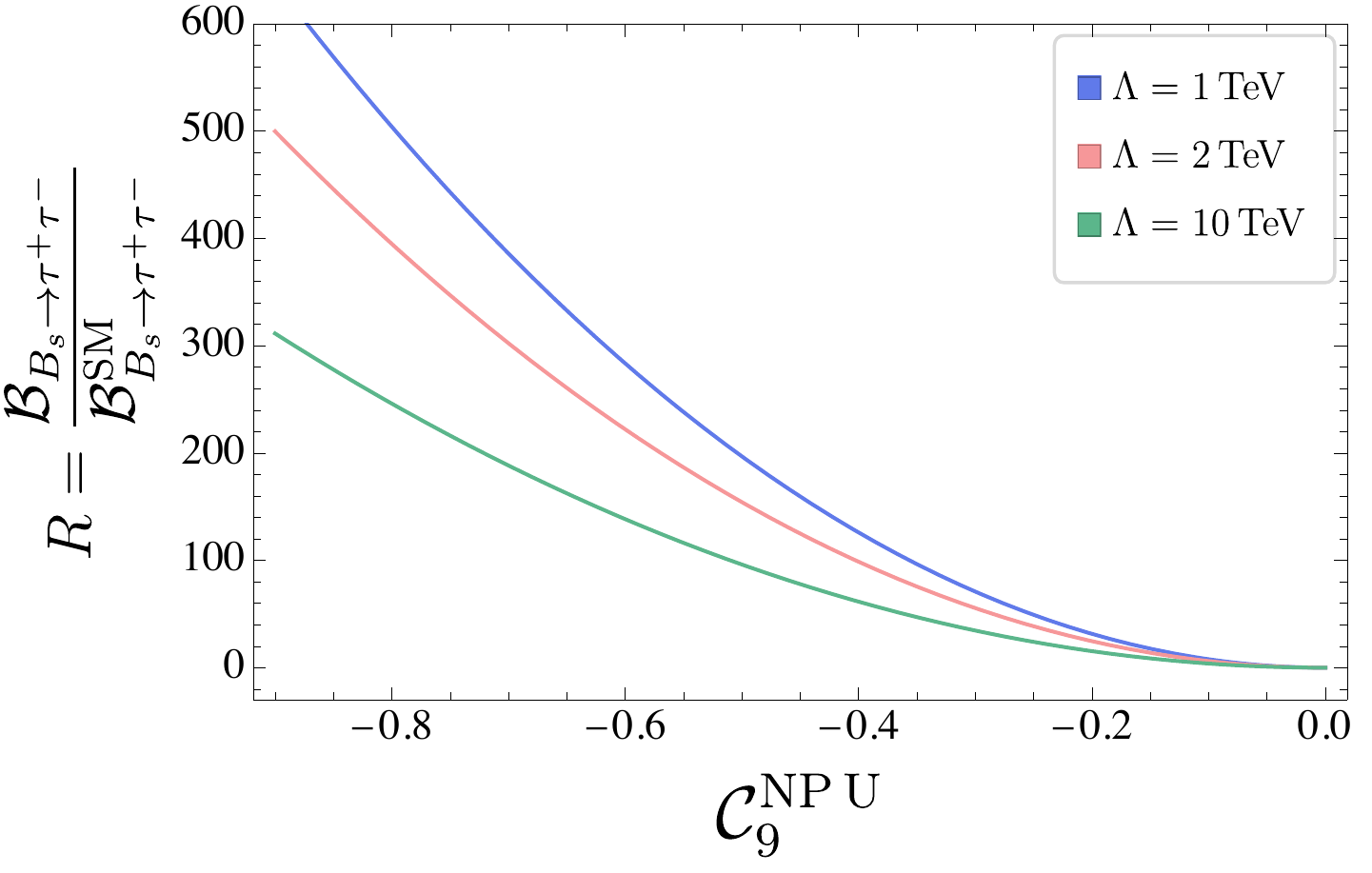}~~~~
    \includegraphics[width=0.435\textwidth]{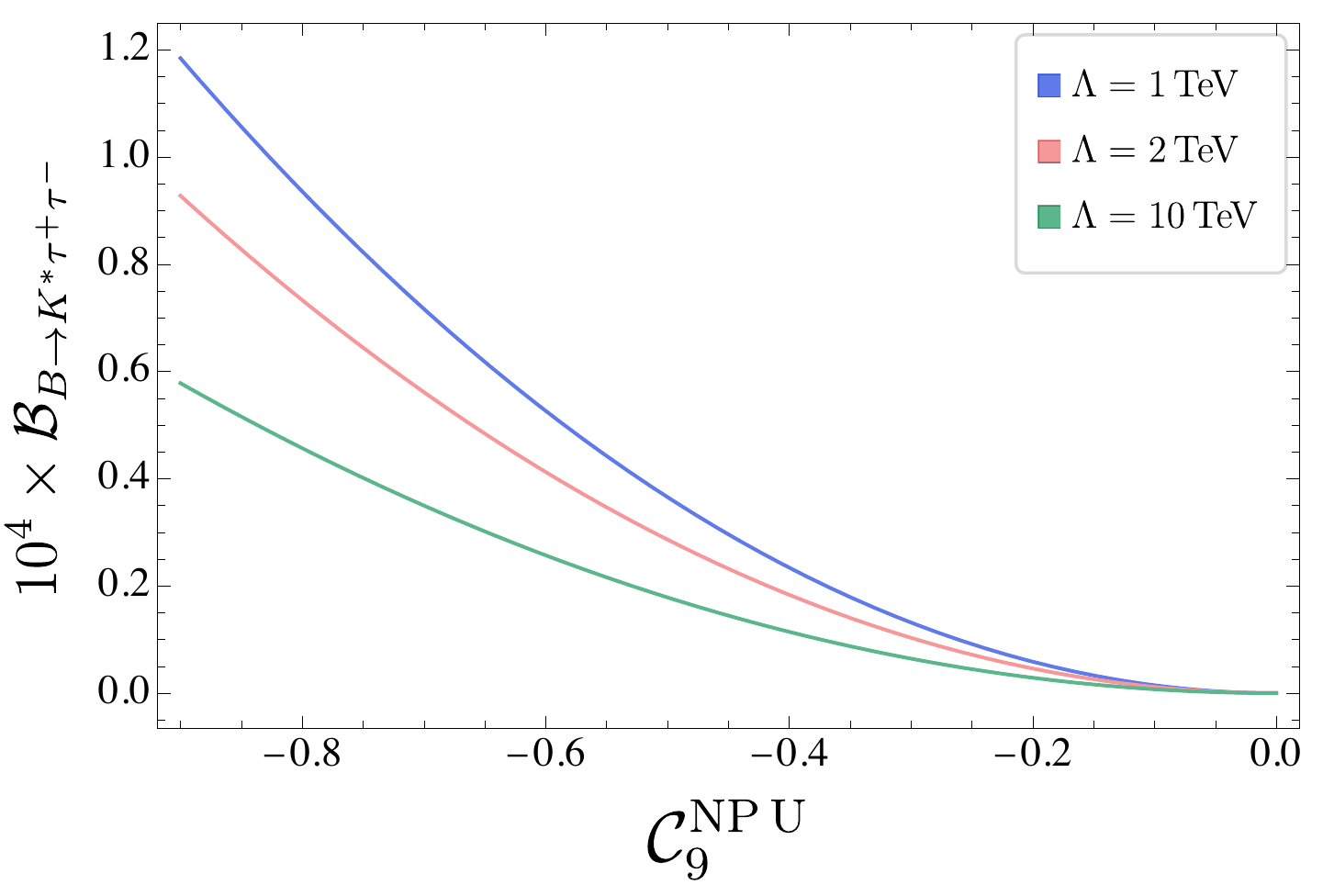}
\caption{Left: Branching ratio of $B_s \to \tau^+\tau^-$ normalized to its SM value as a function of ${\cal C}_9^{\rm NP\, U}$ for different values of the NP scale $\Lambda$ (blue $\Lambda=\SI{1}{TeV}$, orange $\Lambda=\SI{2}{\TeV}$ and green $\Lambda=\SI{10}{TeV}$). Right: Branching ratio of $B \to K^*\tau^+\tau^-$ as a function of ${\cal C}_9^{\rm NP\, U}$. 
A similar plot can be made for $B \to K\tau^+\tau^-$ using Eq.~\eqref{ap3}.
}
\label{fig:C9U}
\end{figure*}

\subsection{Direct NP contribution}

NP above the EW scale is in general chiral (i.e.~it couples to left-handed or right-handed SM fermions). Therefore, a single new particle (with the exception of a $Z^\prime$) does not generate solely a contribution to ${\cal C}_{9}$. Furthermore, in models containing leptoquarks (LQs), or new scalars and fermions contributing at the loop level, in order to get a LFU effect, several generations of such new particles are necessary to avoid the stringent bounds from charged lepton flavour violating processes generated otherwise such as $\mu\to e\gamma$~\cite{Arnan:2016cpy,Crivellin:2017dsk,Crivellin:2022mff}. Therefore, a $Z^\prime$ boson with vectorial leptons couplings seems to be the most natural possibility to generate ${\cal C}_{9}^{\rm NP\, U}$, in particular since such couplings allow for simple gauge anomaly free charge assignments. For example, $3L_\mu+L_e-4L_\tau$ gives a (partially) LFU effect, it can describe data very well and the necessary couplings to $b-s$ can be induced by vector-like quarks~\cite{Altmannshofer:2014cfa,Crivellin:2015mga}.

\subsection{Tau Loops}

Four-fermion operators of the type $\bar s\gamma^\mu P_L b \bar f\gamma_\mu f$, with $f$ being any light SM fermion, can naturally generate ${\cal C}_{9}^{\rm NP \, U}$ via an off-shell photon penguin leading to mixing into ${\cal O}_{9}^{\rm U}$~\cite{Bobeth:2011st}. However, if $f=u,d$, LHC bounds on di-jet searches~\cite{CMS:2021ctt,ATLAS:2020yat} rule out such a possibility. Furthermore, for $f=s,c$ a sizable effect is disfavoured by these constraints, while for $f=e,\mu$ one generates direct LFUV effects dominant over the mixing induced one. Therefore, we are left with $f=b,\tau$. Given that any tree-level NP model generating $\bar s\gamma^\mu P_L b \bar b\gamma_\mu b$ also gives rise to $B_s-\bar B_s$ mixing,  the possible size of this contribution is significantly limited. Therefore, the most unconstrained scenario capable of providing the largest effect in ${\cal C}_{9}^{\rm NP \, U}$ are tau leptons.
In fact, one can find the following model-independent relation
\begin{equation}
{\cal C}_{9}^{\rm NP \, U} =  - \frac{\alpha}{{3\pi }}\log \left( {\frac{{{\Lambda^2}}}{{\mu _b^2}}} \right){\cal C}_{9}^{\tau \tau }\,,
\label{C9UC9tautau}
\end{equation}
originating from the mixing of ${\cal O}_9^{\tau\tau}$ into ${\cal O}_9^{\rm U}$ with $\Lambda$ being the NP scale. Again, since beyond the SM physics is chiral,  at the scale $\Lambda$ also ${\cal C}_{10}^{\tau\tau}$ is in general present. In fact, because leptoquarks are the only particles that can generate a large ${\cal C}_{9}^{\tau\tau}$ without being in conflict with other processes (like $B_s-\bar B_s$ mixing and/or LHC searches), we have either ${\cal C}_{9}^{\tau\tau}={\cal C}_{10}^{\tau\tau}$ in case of the $S_2$ LQ~\cite{Crivellin:2022mff} or ${\cal C}_{9}^{\tau\tau}=-{\cal C}_{10}^{\tau\tau}$ for $U_1$~\cite{Crivellin:2018yvo} of $S_1+S_3$~\cite{Crivellin:2019dwb} unless several representations are combined.\footnote{Note that $U_3$ alone does not work as it induces dangerously large effects in $B\to K^{*}\nu\nu$.} 

Therefore, Eq.~\eqref{C9UC9tautau} allows for a direct correlation between $b\to s\tau^+\tau^-$ processes and ${\cal C}_{9}^{\rm NP \, U}$ if we assume the dominance of ${\cal C}_{9}^{\tau\tau}=\pm {\cal C}_{10}^{\tau\tau}$ at the matching scale and neglect the SM contribution to $b\to s\tau^+\tau^-$ processes. Using
\begin{equation}
\begin{aligned}
\label{ap3}
{\cal B}_{B_s\to\tau\tau}&\approx {\cal B}^{\rm SM}_{B_s\to\tau\tau} \frac{|\Cc{10}^{\tau\tau}|^2}{|\Cc{10}^{\rm SM}|^2}\,,\\[1.5ex]
{\cal B}_{B\to K^*\tau\tau}&\approx   {10^{ - 9}}\times\left(8|\Cc{9}^{\tau\tau}|^2+2|\Cc{10}^{\tau\tau}|^2\right),\,\\[2ex]
{\cal B}_{B\to K\tau\tau}&\approx   {10^{ - 9}}\times\left(3|\Cc{9}^{\tau\tau}|^2+6|\Cc{10}^{\tau\tau}|^2\right),
\end{aligned}
\end{equation}
we show the relation between ${\cal C}_{9}^{\rm NP \, U}$ and these decays in Fig.~\ref{fig:C9U}. One can see that in order for ${\cal C}_9^{\rm NP \, U}$ to be sizable, these processes are significantly enhanced (by orders of magnitude compared to the SM) such that they are within the reach of LHCb and Belle~II. Therefore, a measurement of ${\cal B}_{B_s \to \tau^+\tau^-}$ and/or ${\cal B}_{B \to K^{(*)}\tau^+\tau^-}$ will give us information on the size of $\Cc{9}^{\rm NP \, U}$ (assuming the absence of scalar currents and RHCs). 

\begin{table*}[!ht] 
    \centering
\begin{tabular}{|c||c|c|} 
\hline
 Name of Hypothesis  & Definition & Best fit \\
\hline\hline
Hypothesis I & $\{\Cc{9\mu}^{\rm NP}=-\Cc{9^\prime\mu} , \Cc{10\mu}^{\rm NP}=\Cc{10^\prime\mu}\}$  & (-1.01, +0.31) \\
Hypothesis V or Scenario-R  & $\{\Cc{9\mu}^{\rm NP} , \Cc{9^\prime\mu}=-\Cc{10^\prime\mu}\}$ & (-1.15, +0.17)  \\
\hline
Scenario 8 or Scenario-U & $\{\Cc{9\mu}^{\rm V}=- \Cc{10\mu}^{\rm V}, \Cc{9}^{\rm U}\}$ & $(-0.34,-0.82)$   \\
\hline
Scenario 6 & $\{\Cc{9\mu}^{\rm V}=- \Cc{10\mu}^{\rm V}, \Cc{9}^{\rm U}=\Cc{10}^{\rm U}\}$ & $(-0.52,-0.38)$    \\
Scenario 10 & $\{\Cc{9\mu}^{\rm V}, \Cc{10}^{\rm U}\}$ & $(-0.98,+0.27)$    \\
Scenario 11 & $\{\Cc{9\mu}^{\rm V}, \Cc{10^\prime}^{\rm U}\}$ & $(-1.06,-0.23)$    \\
Scenario 13 & $\{\Cc{9\mu}^{\rm V},\Cc{9^\prime\mu}^{\rm V},\Cc{10}^{\rm U} ,\Cc{10^\prime}^{\rm U}\}$ & $(-1.11,+0.37,+0.28,+0.03)$    \\
\hline
\end{tabular}
\caption{Definition of some of the most prominent scenarios, introduced in Refs.~\cite{Capdevila:2017bsm, Alguero:2018nvb} and updated in Ref.~\cite{Alguero:2021anc}, as well as their best fit points.}
\label{tabdef}
\end{table*}

\subsection{Connection to \texorpdfstring{$R_{D^{(*)}}$}{RDs}}

\begin{figure}[h]
    \centering
    \includegraphics[width=8cm]{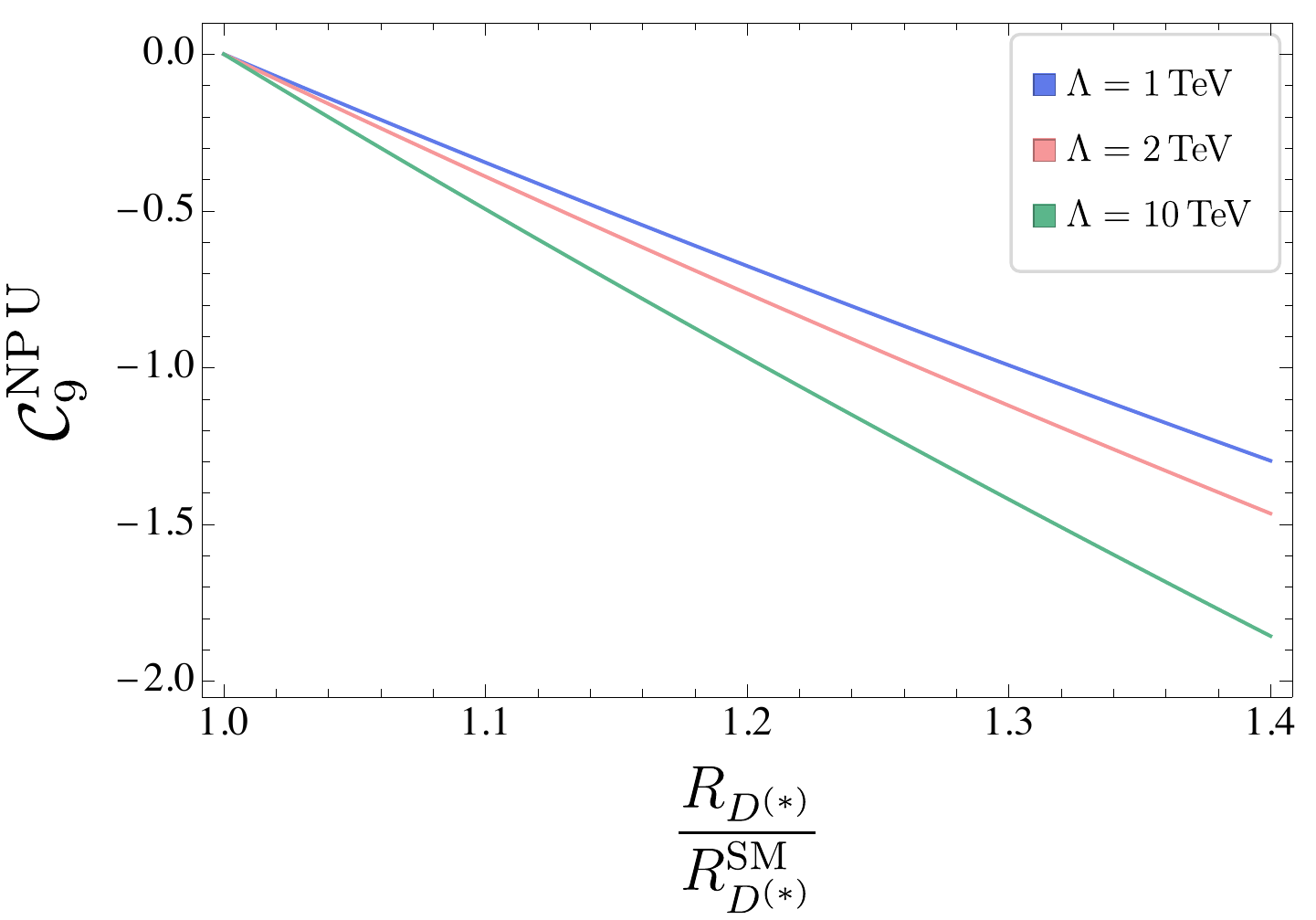}
\caption{${\cal C}_9^{\rm NP\, U}$  versus $R_{D^{(*)}}/R_{D^{(*)}}^{\rm SM}$ for different values of the NP scale $\Lambda$ (blue $\Lambda=\SI{1}{TeV}$, orange $\Lambda=\SI{2}{TeV}$ and green $\Lambda=\SI{10}{TeV}$). }
\label{C9UvsRX}
\end{figure}

Furthermore, in case of ${\cal C}_{9}^{\tau\tau}=-{\cal C}_{10}^{\tau\tau}$, i.e.~if NP is left-handed, a model independent connection within the SMEFT formalism, based on the leptoquark analysis of Ref.~\cite{Crivellin:2018yvo}, between the anomalies in $R_{D^{(*)}}$ and a tau-loop contribution to $b\to s\ell^+\ell^-$ can be established~\cite{Capdevila:2017iqn}: Requiring the absence of dangerously large effects in $b\to s\nu\nu$, and given a generic flavour structure, we have\footnote{In case of partial alignment, i.e. a non-generic flavour structure~\cite{Bordone:2018nbg}, the right-handed side of Eq.~(\ref{C9Utautau}) should be multiplied by a factor $1/(1+{V_{cb}C^{(1,3)}_{3333}/C^{(1,3)}_{2333}})$.}
\begin{equation}
{\cal C}_9^{\tau\tau} =  - {\cal C}_{10}^{\tau \tau } \approx \frac{{ - 2{V_{cb}}}}{{{V_{tb}}V_{ts}^*}}\frac{\pi }{\alpha }\left({\sqrt {{X_{{D^{\left( * \right)}}}}}  - 1}\right) \label{C9Utautau}\,,
\end{equation}
where 
 $X_{D^{(*)}}=R_{D^{(*)}}/R^{\rm SM}_{D^{(*)}}$. We assumed a real NP effect in $R_{D^{(*)}}$, i.e.~with the same phase as the SM contribution. Using \eq{C9UC9tautau}, this translates into
\begin{equation}
\label{ap1}
\Cc{9}^{\rm NP \, U}= \frac{2}{3} \frac{V_{cb}}{V_{tb} V_{ts}^*} \ln\left(\frac{\Lambda^2}{\mu_b^2}\right) \left(\sqrt{X_{D^{(*)}}}-1\right)\,,
\end{equation}
which implies a link between ${\cal C}_{9}^{\rm NP \, U}$ and $X_{D^{(*)}}$, under the given assumptions, as shown in Fig.~\ref{C9UvsRX}\footnote{Note that ${\cal C}_9^{\tau\tau}$ also has a small $q^2$ dependent contribution to $B\to K^{(*)}\ell^+\ell^-$ matrix elements~\cite{Cornella:2020aoq}.}. 

\subsection{Unknown hadronic contribution}

The resulting amount of $\Cc{9}^{\rm NP \, U}$ generated via a direct NP contribution,  tau-loops or all other possible NP sources informs us also of the maximal size of any hypothetical extra hadronic contribution that could have been hidden inside $\Cc{9}^{\rm U}$. This contribution is 
given by the difference between $\Cc{9}^{\rm U}$ and $\Cc{9}^{\rm NP \,  U}$.

In order to get now a rough idea of the size of such hypothetical hadronic quantity that we will refer as $\Delta$ and given the current absence of a precise measurement of $Q_5$ from LHCb, we can proceed  in the following way. Let's assume that Scenario 8 or Scenario-U (see Table~\ref{tabdef} for definition), which predicts a sizable $\Cc{9}^{\rm U}$ and currently gives the best fit to data of all scenarios is the right scenario implemented in Nature. Assuming in addition the link with $R_{D^{(*)}}$, the difference between $\Cc{9}^{\rm U}$ obtained from the global fit (see Table~\ref{tabdef}), without using the connection with $R_{D^{(*)}}$, and $\Cc{9}^{\rm NP\, U}$ obtained using Eq.~\eqref{ap1} for $\Lambda=\SI{2}{TeV}$ gives us an idea of an upper bound on an  unknown hadronic contribution. With present data, in Scenario-U this amounts roughly to a value for $\Delta\sim -0.28$, which is a rather small effect ($\approx6.8\%$) compared to the SM value of $\Cc{9\mu}$, especially considering that the NP contribution to this coefficient is of order $20\%$ of its SM value. But more interesting is the fact that this quantity depends on the NP scale ($\Lambda$)\footnote{In order to keep couplings in a perturbative regime, the scale of NP cannot go beyond $\SI{10}{TeV}$.}. This implies that a larger NP scale ($\Lambda$) leads also to a larger ${\cal C}_9^{\rm NP \, U}$  (given a fixed $R_{D^{(*)}}$) and in turn $\Delta$ becomes even smaller (in absolute value) up to $\Delta\sim -0.13$ for $\Lambda=\SI{10}{TeV}$. 

\begin{figure}[h]
    \centering
        \includegraphics[width=0.48\textwidth]{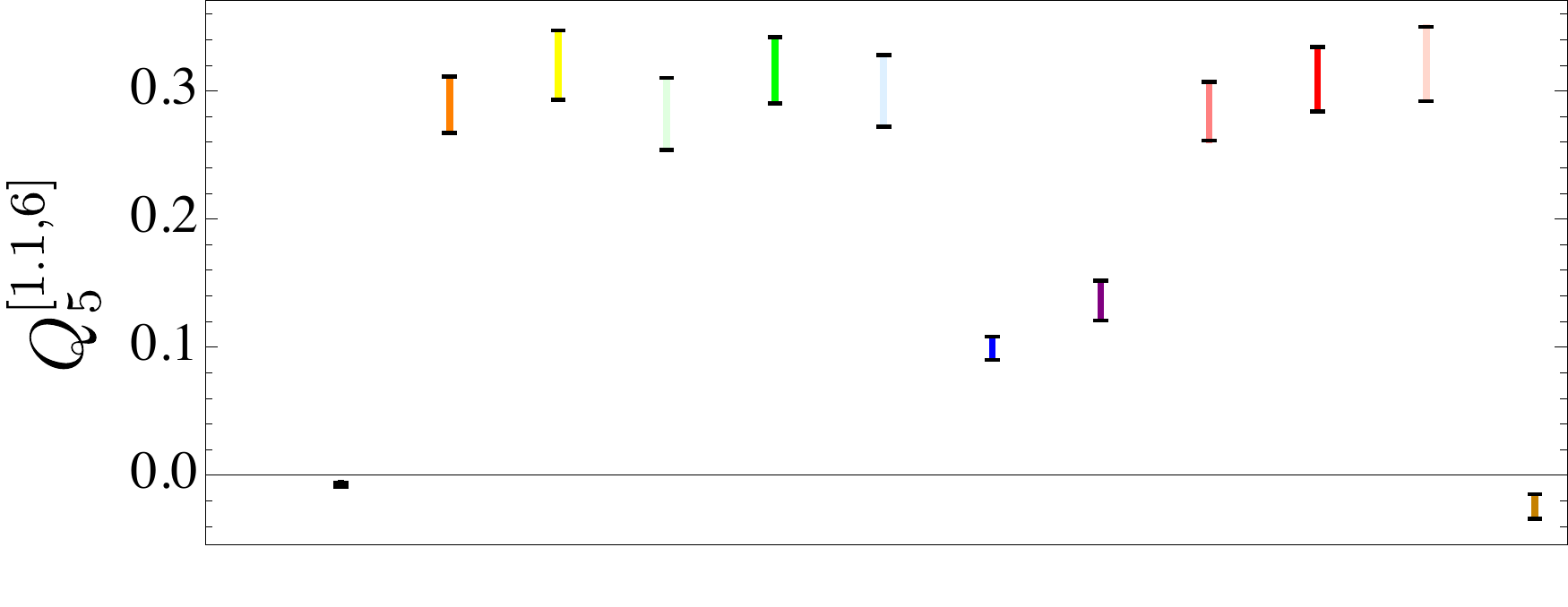}
\caption{$Q_5^{[1.1,6]}$ for different scenarios. From left to right: SM, ${\cal C}_{9\mu}^{\rm NP}$, $\{{\cal C}_{9\mu}^{\rm NP}, {\cal C}_{9'\mu}\}$, $\{{\cal C}_{9\mu}^{\rm NP}, {\cal C}_{10'\mu}\}$, Hyp~I, Hyp~V, Sc.~8, Sc.~6, Sc.~10, Sc.~11, Sc.~13 and ${\cal C}^{\rm NP}_{10\mu}$. See Table~\ref{tabdef} for explicit definitions of hypotheses and scenarios in terms of Wilson coefficients. The scenario ${\cal C}^{\rm NP}_{10\mu}$ is included as an example of a non-preferred NP scenario (smaller Pull$_{\rm SM}$) that can be easily disfavoured with a precise measurement of $Q_5^{[1.1,6]}$. }
\label{fig:plot3}
\end{figure}

\section{What do we learn from \texorpdfstring{$Q_5$}{Q5}?}
\label{sec:NPQ5}

As we have discussed in the previous section, $Q_5$ is a unique discriminator of the presence of a LFUV piece in ${\cal C}_{9\mu}^{\rm NP}$. So far, only Belle has provided a measurement of this observable~\cite{Belle:2016fev} (using an average of neutral and charged decays to increase the statistics), albeit with very large error bars,
\begin{equation}
    Q_5^{[1.1,6]}=0.656\pm 0.485 \pm 0.103\,.
\end{equation}

The ten most strongly preferred NP scenarios, i.e.~with Pull$_{\rm SM}$ above 6.5$\sigma$, (see Fig.~\ref{fig:plot3}) can be classified in two groups:
\begin{itemize}
    \item Scenarios with ${\cal C}_{9\mu}^{\rm V} \lesssim -1$: ${\cal C}_{9\mu}^{\rm NP}$, $\{{\cal C}_{9\mu}^{\rm NP}, {\cal C}_{9^\prime\mu}\}$,   $\{{\cal C}_{9\mu}^{\rm NP}, {\cal C}_{10^\prime\mu}\}$, Hypotheses I and V (or Scenario-R), Scenarios 10, 11, 13 (see Table~\ref{tabdef} for definitions). They all predict $Q_5 \approx +0.3$. The largest Pull$_{\rm SM}$ was found in Ref.~\cite{Alguero:2021anc} for Scenario-R (see Table~\ref{tabdef}).
    In this scenario the contribution of ${\cal C}_{9\mu}^{\rm NP}$ is counterbalanced by the contribution of the primed coefficients (RHCs) in observables like $R_K$.
    \item Scenarios where the contribution to ${\cal C}_{9\mu}^{\rm NP}$ contains both a LFUV and a LFU piece with the absolute value of the former being smaller compared to the previous case: Scenario-U and Scenario 6. The reason why the absolute value of the LFUV piece is smaller than in the previous case is the link to ${\cal C}_{10\mu}^{\rm NP}$, resulting in a bound from ${\cal B}_{B_s \to \mu^+\mu^-}$. The largest Pull$_{\rm SM}$ is found for Scenario-U (also called Scenario 8) predicting $Q_5 \approx +0.1$. 
\end{itemize}

This clearly indicates that while $R_K$ is fundamental to establish the violation of LFU, finding $Q_5$ close to its SM value would by no means be contradictory to the violation of LFU in $R_K$. In other words, it is essential to consider the pair $R_K$ and $Q_5$ (but also other $R_{X}$ observables). Therefore, a measurement of $Q_5$ close to $+0.3$ would point to one of the scenarios of the first group (including those with RHCs), while a $Q_5\sim +0.1$ (very close to SM) would indicate the presence of LFU NP. No other observable has this power to discriminate among these two kinds of scenarios.

\begin{figure}[t!]
\centering
    \includegraphics[width=0.54\textwidth,height=8cm]{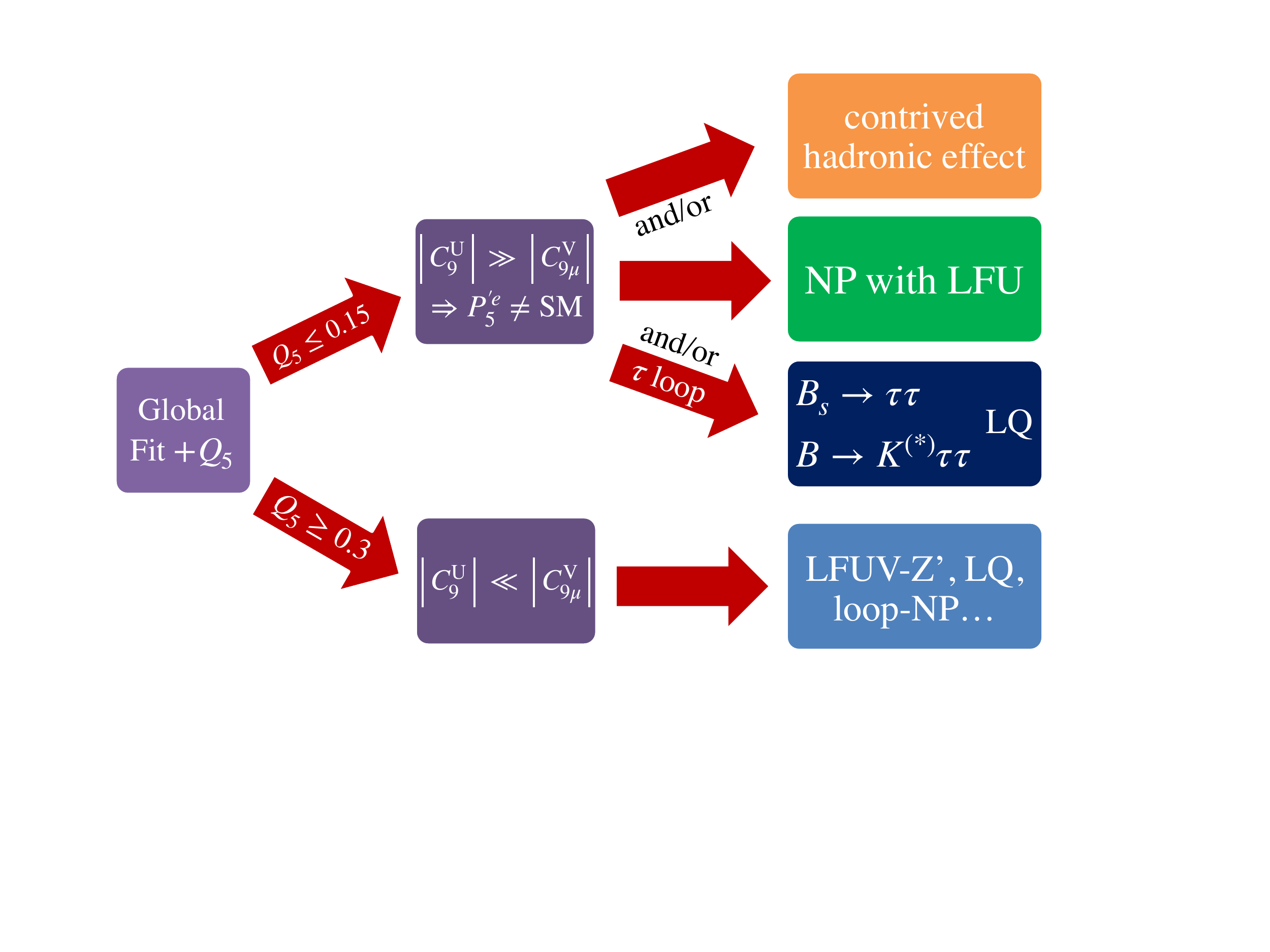}
\caption{Decision tree depending on the possible values of $Q_5$ and future measurements of $b \to s \tau^+\tau^-$ decays.}
\label{fig:decisiontree}
\end{figure}

We observed no difference (in terms of Pull$_{\rm SM}$) between Scenario-R and Scenario-U along the years (in 2019, 2020, 2021), with $7.1\sigma$ (Scenario-R) and $7.2\sigma$ (Scenario-U) in the latest analysis. Here, we have studied the impact of a measurement of $Q_5^{[1.1,6]}$ using all data entering the global fits in Table~\ref{tab:results1D_a},  assuming two experimental values for this observable: $0.30\pm 0.10$ and $0.10\pm 0.10$. If the measurement falls within the former range no disentangling is possible, while for the latter some marginal disentangling arises. On the other hand, if the precision on $Q_5^{[1.1,6]}$ improves by a factor of 2 (see Table~\ref{tab:results1D_b}), some difference of around $0.7\sigma$ appears between the two options. The corresponding results using only LFUV observables are also displayed in Table~\ref{tab:results1D_a} and Table~\ref{tab:results1D_b}.

\begin{table*}[t] 
    \centering
\begin{tabular}{c||c|c|c||c|c|c} 
($Q_5=0.3\pm 0.1$) & \multicolumn{3}{c||}{All} &  \multicolumn{3}{c}{LFUV}\\
\hline
Best Scenarios   & Best fit 
& Pull$_{\rm SM}$ & p-value & Best fit 
& Pull$_{\rm SM}$ & p-value\\
\hline\hline
\multirow{2}{*}{$\{\Cc{9\mu}^{\rm V}=-\Cc{10\mu}^{\rm V},\Cc{9}^{\rm U}\}$}    & \multirow{2}{*}{[-0.34,-0.82]} &   
\multirow{2}{*}{7.4}   & \multirow{2}{*}{31.8\,\%}
&   \multirow{2}{*}{[-0.39,-1.86]}   
&
\multirow{2}{*}{5.2}  & \multirow{2}{*}{46.4\,\%}  \\
 &  &   &  &   & \\
 \multirow{2}{*}{$\{\Cc{9\mu}^{\rm NP}, \Cc{9^\prime\mu}=-\Cc{10^\prime\mu}\}$}    &   \multirow{2}{*}{[-1.15,+0.22]} &   
 \multirow{2}{*}{7.5}  & \multirow{2}{*}{35.6\,\%}
 &  \multirow{2}{*}{[-1.45,+0.33]}   
 & \multirow{2}{*}{5.7}   & \multirow{2}{*}{77.9\,\%}  \\
 &  &  & &  &  
\end{tabular} 
\begin{tabular}{c||c|c|c||c|c|c} 
($Q_5=0.1\pm 0.1$)  & \multicolumn{3}{c||}{All} &  \multicolumn{3}{c}{LFUV}\\
\hline
Best Scenarios   & Best fit  
& Pull$_{\rm SM}$ & p-value & Best fit 
& Pull$_{\rm SM}$ & p-value\\
\hline\hline
\multirow{2}{*}{$\{\Cc{9\mu}^{\rm V}=-\Cc{10\mu}^{\rm V},\Cc{9}^{\rm U}\}$}    & \multirow{2}{*}{[-0.31,-0.83]} &   
\multirow{2}{*}{7.1}   & \multirow{2}{*}{38.0\,\%}
&   \multirow{2}{*}{[-0.40,+0.15]}   
&\multirow{2}{*}{4.7}  & \multirow{2}{*}{66.7\,\%}  \\
 &  &   &  &  & \\
 \multirow{2}{*}{$\{\Cc{9\mu}^{\rm NP}, \Cc{9^\prime\mu}=-\Cc{10^\prime\mu}\}$}    &   \multirow{2}{*}{[-1.04,+0.20]} &   
 \multirow{2}{*}{6.7}  & \multirow{2}{*}{30.1\,\%}
 &  \multirow{2}{*}{[-1.08,+0.23]}   
 & \multirow{2}{*}{4.5}   & \multirow{2}{*}{49.8\,\%}  \\
 &   &  & &  &  
\end{tabular} 
\caption{Fit results for two different scenarios using current data and in addition assuming $Q_5^{[1.1,6]}=0.30\pm 0.10$ (upper part) and  $Q_5^{[1.1,6]}=0.10\pm 0.10$ (lower part). }
\label{tab:results1D_a}
\end{table*}
\begin{table*}[t] 
    \centering
\begin{tabular}{c||c|c|c||c|c|c} 
($Q_5=0.30\pm 0.05$) & \multicolumn{3}{c||}{All} &  \multicolumn{3}{c}{LFUV}\\
\hline
Best Scenarios   & Best fit 
& Pull$_{\rm SM}$ & p-value & Best fit 
& Pull$_{\rm SM}$ & p-value\\
\hline\hline
\multirow{2}{*}{$\{\Cc{9\mu}^{\rm V}=-\Cc{10\mu}^{\rm V},\Cc{9}^{\rm U}\}$}    & \multirow{2}{*}{[-0.42,-0.79]} &   
\multirow{2}{*}{8.5}   & \multirow{2}{*}{19\,\%}
&   \multirow{2}{*}{[-0.49,-2.61]}   
&
\multirow{2}{*}{7.0}  & \multirow{2}{*}{19.2\,\%}  \\
 &  &  
 &  & 
 & \\
 \multirow{2}{*}{$\{\Cc{9\mu}^{\rm NP}, \Cc{9^\prime\mu}=-\Cc{10^\prime\mu}\}$}    &   \multirow{2}{*}{[-1.15,+0.21]} &   
 \multirow{2}{*}{9.2}  & \multirow{2}{*}{35.6\,\%}
 &  \multirow{2}{*}{[-1.29,+0.29]}   
 & \multirow{2}{*}{7.7}   & \multirow{2}{*}{74.5\,\%}  \\
 & 
 &  & & 
 &   
\end{tabular} 
\begin{tabular}{c||c|c|c||c|c|c} 
($Q_5=0.10\pm 0.05$)  & \multicolumn{3}{c||}{All} &  \multicolumn{3}{c}{LFUV}\\
\hline
Best Scenarios   & Best fit  
& Pull$_{\rm SM}$ & p-value & Best fit 

& Pull$_{\rm SM}$ & p-value\\
\hline\hline
\multirow{2}{*}{$\{\Cc{9\mu}^{\rm V}=-\Cc{10\mu}^{\rm V},\Cc{9}^{\rm U}\}$}    & \multirow{2}{*}{[-0.30,-0.84]} &   
\multirow{2}{*}{7.3}   & \multirow{2}{*}{38.0\,\%}
&   \multirow{2}{*}{[-0.40,-0.07]}   
&
\multirow{2}{*}{5.1}  & \multirow{2}{*}{66.6\,\%}  \\
 & 
 &  
 &  &  
 & \\
 \multirow{2}{*}{$\{\Cc{9\mu}^{\rm NP}, \Cc{9^\prime\mu}=-\Cc{10^\prime\mu}\}$}    &   \multirow{2}{*}{[-0.85,+0.18]} &   
 \multirow{2}{*}{6.6}  & \multirow{2}{*}{22.2\,\%}
 &  \multirow{2}{*}{[-0.76,+0.16]}   
 & \multirow{2}{*}{4.4}   & \multirow{2}{*}{30.2\,\%}  \\
 & 
 &  & & 
 &   
\end{tabular} 
\caption{Fit results for two different scenarios using current data and assuming in addition $Q_5^{[1.1,6]}=0.30\pm 0.05$ (upper part) and  $Q_5^{[1.1,6]}=0.10\pm 0.05$ (lower part).}
\label{tab:results1D_b}
\end{table*}

\section{Conclusions and Outlook}
\label{sec:conclusions}

In summary, a combined measurement of $P_{5}^{\prime\mu}$, $Q_5$, $R_{X}$ (with $X=K,K^*,K_S,K^{*+},\phi$)
 and $b\to s\tau^+\tau^-$ processes (once all data is available), together with $R_{D^{(*)}}/R^{\rm SM}_{D^{(*)}}$, can provide us not only with a clear signal of NP, but also will inform us about its structure. In particular, this will tell us the size of ${\cal C}_{9}^{\rm U}$ and can possibly even uncover the origin of this Wilson coefficient. In this context, $Q_5$ is of utmost importance to make progress by breaking the present degeneracy among the scenarios that currently explain data best.
 
Each of those observables plays a specific but complementary role: $R_{X}$ observables determine the breaking of LFU, $P_5^{\prime \mu}$ gives the size of NP (and possible unknown hadronic contributions) within ${\cal C}_{9 \mu}$, $Q_5$ determines the amount of LFUV in ${\cal C}_{9\mu}^{\rm NP}$. ${\cal B}_{B_s \to \tau^+\tau^-}$, if found to be very large, is a clear NP signal, and if linked to $R_{D^{(*)}}/R^{\rm SM}_{D^{(*)}}$, determines the amount of LFU NP in ${\cal C}_{9\mu}^{\rm NP}$.
In turn a large LFU NP implies 
  the smallness of any unknown hadronic contribution (beyond those already included in our conservative setup outlined in Sec.~\ref{sec:setup}). The resulting proposed procedure for determining the origin of NP is shown graphically in Fig.~\ref{fig:decisiontree}.

\begin{acknowledgements}
	The work of A.C.~is supported by a Professorship Grant (PP00P2\_176884) of the Swiss National Science Foundation. J.M.~ gratefully acknowledges the financial support by ICREA under the ICREA Academia programme. JM and MA received financial support from Spanish Ministry of Science, Innovation and Universities (project PID2020-112965GB-I00) and from the Research Grant Agency of the Government of Catalonia (project SGR 1069). B.C. is supported by the Italian Ministry of Research (MIUR) under the Grant No. PRIN 20172LNEEZ. 
\end{acknowledgements}

\section{Appendix}
In Table~\ref{tab:results1D_a} and Table~\ref{tab:results1D_b} we provide the prediction for the best scenarios (highest Pull$_{\rm SM}$) of the global fits using all 254 observables plus $Q_5$, as well as the fit including only LFUV observables for a small and a sizable value of $Q_5^{[1.1,6]}$ assuming an error of 0.10 and 0.05, respectively.

\bibliography{bibliography}

\end{document}